\documentclass[12pt]{article} 
\usepackage{amsmath}
\usepackage{graphics}
\usepackage{feynmp}
\catcode`\@=11 
 
\def\mycite{\@ifnextchar [{\@tempswatrue\@mycitex}{\@tempswafalse\@mycitex[]}} 
\def\mcite{\@ifnextchar [{\@tempswatrue\@mycitex}{\@tempswafalse\@mycitex[]}} 
 
\def\@mycitex[#1]#2{\if@filesw\immediate\write\@auxout{\string\citation{#2}}\fi 
 \def\@citea{}\@mycite{\@for\@citeb:=#2\do 
    {\@citea\def\@citea{,\penalty\@m\ }\@ifundefined 
       {b@\@citeb}{{\bf ?}\@warning 
       {Citation `\@citeb' on page \thepage \space undefined}}%
\hbox{\csname b@\@citeb\endcsname}}}{#1}}

\def\@mycite#1{[{#1}]} 
 
\catcode`\@=12 
\def\e{{\, e}\,}
\newcommand{\rf}[1]{(\ref{#1})}

\newcommand{\be}{\begin{equation}}
\newcommand{\ee}{\end{equation}}
\def\bea{\begin{eqnarray}}
\def\eea{\end{eqnarray}}

\newcommand\N{${\cal N}=4~$}
\newcommand\Tr{\mathop{\mathrm{Tr}}}
\def\a{\alpha}
\def\l{\lambda}

\begin{document} 
\begin{titlepage} 
\begin{flushright} 
NBI--HE--99--45\\ 
ITEP--TH--60/99 \\
hep-th/9911088\\
\hfill{ }\\
November 1999
\end{flushright} 
\vspace{0.5cm} 
 
\begin{center} 
{\large Static Potential in ${\cal N}=4$ Supersymmetric Yang-Mills Theory}\\ 
\vspace{1cm} 
{ J.K.~Erickson$^1$, 
G.W.~Semenoff$\,^{1}$,
R.J.~Szabo$\,^{3}$ 
and K.~Zarembo$^{1,2,4}$}
\hfill\vskip 5pt 
\hspace*{0.37em}E-mail: \texttt{erickson@nbi.dk}, 
\texttt{semenoff@nbi.dk}, \texttt{szabo@nbi.dk},
\texttt{zarembo@physics.ubc.ca} 
\\ 
\vspace{24pt}
${}^1${\it Department of Physics and Astronomy}\\ 
${}^2${\it Pacific Institute for the Mathematical Sciences
\\University of British Columbia 
\\Vancouver, British Columbia V6T 1Z1, Canada} 
\vskip 0.5cm
${}^3${\it The Niels Bohr Institute\\ 
Blegdamsvej 17\\ 
DK-2100 Copenhagen \O, 
Denmark}
\vskip 0.5cm
${}^4${\it Institute of Theoretical and Experimental  
Physics\\ B. Cheremushkinskaya 25,
Moscow, 117218 Russian Federation}
\end{center} 
\vskip 1 cm 
\begin{abstract} 
  We compute the leading order perturbative correction to the static
  potential in ${\cal N}=4$ supersymmetric Yang-Mills theory.  We show
  that the perturbative expansion contains infrared logarithms which,
  when resummed, become logarithms of the coupling constant.  The
  resulting correction goes in the right direction to match the strong
  coupling behavior obtained from the AdS/CFT correspondence.  We find
  that the strong coupling extrapolation of the sum of ladder diagrams
  goes as $\sqrt{g^2N}$, as in the supergravity approach.
\end{abstract}
\end{titlepage}
\newpage\setcounter{page}{1}

\setcounter{equation}{0} 

\setcounter{page}{2}
\begin{fmffile}{fmfstatic}

Yang-Mills theory with \N supersymmetry in four dimensions is an
extremely interesting quantum field theory.  It is a conformal field
theory for any value of its coupling constant and is conjectured to
have a Montonen-Olive strong-weak coupling duality. It is now believed
to have an additional, exact duality to type IIB superstring theory on
an $AdS_5\times S^5$ background.  This duality, which is known as the
AdS/CFT correspondence \cite{Mal97,Gub98,Wit98,Aha99}, is an explicit
realization of the long anticipated but elusive mapping between gauge
theory and string theory \cite{amp}.  In this duality, the weak
coupling, classical limit of string theory corresponds to the large
$N$ limit of gauge theory.  Classical string theory in its low energy,
weak curvature limit is accurately described by classical IIB
supergravity. This corresponds to the large $N$, large 't~Hooft
coupling (denoted $\l\equiv g^2 N$) limit of \N supersymmetric Yang-Mills
theory.  This gives a new, computationally tractable scheme for
extracting the large $\l$ limit of certain correlators in \N supersymmetric
Yang-Mills theory.

It is difficult to check the results of these computations since the
only other quantitative approach to Yang-Mills theory is its
perturbative expansion in small $g^2$.  An important exception is
correlation functions which are protected from gaining quantum
corrections by non-renormalization theorems. The agreement between
supergravity calculations of these correlators and weak coupling
results is a powerful consistency check of the AdS/CFT correspondence
~\cite{Aha99}. In fact, some surprising new non-renormalization
conjectures were found in this way~\cite{8,dine}.

However, the AdS/CFT correspondence predicts a generically non-trivial
dependence of various quantities on the coupling constant.  The sum
over planar diagrams which emerges in the large $N$ limit of
Yang-Mills theory has a finite radius of convergence~\cite{thooft}.
This suggests that the dependence of various quantities on the
coupling constant is smooth near $\l=0$.  Furthermore, since evidence
from matrix models suggests that the singularity in the sum of planar
graphs usually occurs for {\it negative} $\l$~\cite{bipz}, one might
extrapolate this smooth behavior to the strong coupling limit, so that
some quantities have a smooth interpolation between $\l=0$ and
$\l=\infty$.  An example is the free energy of \N supersymmetric
Yang-Mills theory at finite temperature which, because of conformal
invariance, has the form
\[
F/V=-f(g^2,N)\pi^2N^2T^4/6.
\]
In the large $N$ limit $f$ is a function of the 't~Hooft coupling
$\l$, which, to agree with perturbative computation, must go to 1 at
small $\l$ and, to agree with the AdS/CFT correspondence, must go to
3/4 as $\l$ goes to infinity~\cite{Gub96}.  Smoothness of this
function implies that its perturbative expansion has alternating signs
and that corrections to the supergravity result are positive. Explicit
calculations confirmed that this is indeed the case: for large
$\l$~\cite{Gub98'}
$$f=\frac{3}{4}+\frac{45}{32}\zeta(3)\lambda^{-3/2}+\cdots$$ and, for
small $\l$ \cite{Fot98},
$$f=1-\frac{3\lambda}{2\pi^2}+\frac{(3+\sqrt{2})\l^{3/2}}{\pi^3}+\cdots.$$
The latter expression contains the slight surprise that the final term
is non-analytic in small $\l$.  This non-analyticity comes from
infrared divergences that occur at finite temperature.

It is interesting to ask whether infrared divergences lead to
non-analyticity in other quantities. If we take the duality to the IIB
superstring seriously, this would imply something about the strong
curvature limit of tree level string theory.  Local operators and
extensive variables such as the free energy in Yang-Mills theory are
related to the low-energy supergravity degrees of freedom in string
theory.  An object which couples more directly to stringy degrees of
freedom is the Wilson loop.  The loop operator which is a source for
classical strings is~\cite{Mal98}
\begin{equation}
\label{wl}
W(C)=\frac{1}{N}\,\Tr {{\rm P}}\exp\oint_C d\tau\,\left(
iA_\mu\dot{x}_\mu
+\Phi_i\theta_i\sqrt{\dot{x}^2}\right). 
\end{equation} 
It contains a coupling to the six scalar fields $\Phi_i$ of the \N
supermultiplet (with $\theta_i$, $i=1,\ldots,6$ a point on the
5-sphere, $\theta_i^2=1$) as well as the familiar dependence on the
Yang-Mills field. It is related to the holonomy of the wavefunction
of a heavy particle with the quantum numbers of a W-boson.

The simplest quantity which can be extracted from the Wilson loop is
the interaction potential between static W-bosons. For this, we
consider a rectangular loop of length $T$ and width $L$ and the limit
\begin{equation}
V(L)=-\lim_{T\rightarrow\infty}\,\frac{1}{T}\,\ln \left\langle W(C)
\right\rangle.
\label{statpot}
\end{equation} 
Conformal symmetry of \N supersymmetric Yang-Mills theory implies that
\begin{equation}
V(L)=-\frac{\alpha(g^2,N)}{ L},
\end{equation}
and, in the large $N$ limit, the effective Coulomb charge
$\alpha$ is a function of the 't~Hooft coupling $\lambda$.

The exchange of scalar bosons mediates an attractive force which is
equal to that due to gauge fields, so that, at weak coupling, the
potential is twice as large as that in pure Yang-Mills theory.  At
large $\l$, the AdS/CFT correspondence predicts a square-root
dependence of the Coulomb charge~\cite{Mal98}, so that
\begin{equation}
\a(\l)=
\begin{cases}
{\displaystyle\frac{\l}{4\pi}+\cdots}, & \text{for }\l\rightarrow 0; \\[7pt]
{\displaystyle\frac{4\pi^2\sqrt{2\l}}{\Gamma^4(1/4)}
+O(1)}, & \text{for }\l \rightarrow\infty.
\end{cases}
\label{adscft}
\end{equation}
If $\alpha(\l)$ is a smooth function, we expect that perturbative
corrections should decrease the weak coupling value of the Coulomb
charge.  In this Letter, we shall check this assertion by computing
the correction to $\a(\l)$ to the next order in $\l$. Similar
computations for non-supersymmetric Yang-Mills theory have been done
in~\cite{Fis77,App78}.

Feynman rules for \N supersymmetric Yang-Mills theory follow from the
action 
\[
\begin{split}
S=& \, \frac{1}{g^2}\int
d^4x~\biggl\{\tfrac{1}{4}\,(F_{\mu\nu}^a)^2+\tfrac{1}{2}\,(D_\mu\Phi_i^a)^2
+\tfrac{i}{2}\,\overline{\Psi}^a\,\Gamma^\mu
D_\mu\Psi^a\\
&\qquad
+\tfrac{1}{2}\,f^{abc}\,\overline{\Psi}^a\,
\Gamma^i\Phi_i^b\,\Psi^c
+\tfrac{1}{2}\,f^{abc}f^{ade}\sum_{i<j}
\Phi_i^a\Phi_j^b\Phi_i^d
\Phi_j^e\biggr\}
\end{split}
\]
where $i,j$ range from $1$ to $6$,
$D_\mu(\cdot)^a\equiv\partial_\mu(\cdot)^a+f^{abc}A_\mu^b(\cdot)^c$,
and $f^{abc}$ are the structure constants of the $SU(N)$ Lie algebra
(we use the normalization of generators $\Tr T^aT^b=\delta^{ab}/2$).
The $\Psi^a$ are four-component Majorana fermion fields transforming as a
four dimensional representation of the $SO(6)$ R-symmetry group.  We use
Feynman gauge where the gluon propagator is 
$D_{\mu\nu}(x)=\delta_{\mu\nu}g^2/4\pi^2 x^2$, and consider the
contribution of all planar diagrams to the Wilson loop to order
$g^4$. We have found that individual diagrams contain ultraviolet
divergences, which cancel when all diagrams to order
$g^4N^2$ are summed.  Nevertheless, the sum of all one-loop
contributions to the Wilson loop expectation value does not yield a
well-defined correction to the potential. The planar ladder diagram
contains an extra logarithm of $T/L$:
\begin{equation}
-\ln
\biggl\{
1+
\parbox{8mm}{
\begin{fmfgraph}(8,10) \fmfstraight
        \fmfleft{f1,s1,f2}
        \fmfright{f3,s2,f4}
        \fmf{wiggly}{s1,s2}
        \fmf{plain,width=2}{f1,s1,f2}
        \fmf{plain,width=2}{f3,s2,f4}
\end{fmfgraph}}
+
\parbox{8mm}{
\begin{fmfgraph}(8,10) \fmfstraight
        \fmfleft{f1,s1,t1,f2}
        \fmfright{f3,s2,t2,f4}
        \fmf{wiggly}{s1,s2}
        \fmf{wiggly}{t1,t2}
        \fmf{plain,width=2}{f1,s1,t1,f2}
        \fmf{plain,width=2}{f3,s2,t2,f4}
\end{fmfgraph}}
+
\cdots
\biggr\}
=\frac{\l}{4\pi}\frac{T}{L}-\frac{\l^2}{(2\pi)^3}\,\frac{T}{L}\,\ln\frac{T}{L}+\cdots.
\label{ladders}
\end{equation}
This logarithmic term threatens to ruin perturbation theory as well as
the definition of static potential (\ref{statpot}).  A similar
behavior (beginning at order $g^6$) has been observed for Wilson loops
in ordinary Yang-Mills theory.  According to~\cite{App78} the
logarithm is due to an infrared divergence coming from soft gluons
traveling along the Wilson loop for a long period of (Euclidean) time
$t\propto T$.  Diagrams with more soft gluons will have a higher
degree of divergence so it is necessary to take all orders into
account.  It was argued in \cite{App78} that the effect of the soft
gluon resummation is to cut off the divergent integral over $t$ at
$t\sim 1/\l$. Prototypically, an integral like $\int dt/t$ is replaced
by $\int (dt/t)\exp(-\l t)$. The perturbative expansion of the latter
produces successively worse infrared divergences, but resummation of
perturbation series makes the result finite and replaces $\ln(T/L)$ by
$\ln(1/\l)$.

The consistent resummation prescription which removes the infrared
divergences in the static potential is to sum up all (planar) ladder diagrams.  The
sum of ladder diagrams can be obtained from a Bethe-Salpeter equation,
which in diagrammatic form is
\[
\sum_{\text{ladders}}
\parbox{20mm}{\fmfframe(5,5)(5,5){
\begin{fmfgraph*}(12,18)\fmfstraight
\fmfleft{s0,s,s1,s2,s3,s4,S}
\fmfright{t0,t,t1,t2,t3,t4,T}
\fmf{plain,width=2}{s0,s,s1,s2}
\fmf{dots}{s2,s3}
\fmf{plain,width=2}{s3,s4,S}
\fmf{plain,width=2}{t0,t,t1,t2}
\fmf{dots}{t2,t3}
\fmf{plain,width=2}{t3,t4,T}
\fmf{wiggly}{s,v}
\fmf{wiggly}{s1,v}
\fmf{wiggly,tension=2}{s4,v}
\fmf{wiggly}{t,v}
\fmf{wiggly}{t1,v}
\fmf{wiggly,tension=2}{t4,v}
\fmfv{d.sh=circle,d.f=empty,d.si=20}{v}
\fmflabel{$0$}{s0}
\fmflabel{$0$}{t0}
\fmflabel{$S$}{S}
\fmflabel{$T$}{T}
\end{fmfgraph*}}}
=
1+
\sum_{\text{ladders}}
\parbox{20mm}{\fmfframe(5,5)(5,5){
\begin{fmfgraph*}(12,18)\fmfstraight
\fmfleft{s0,s,s1,s2,s3,s4,s5,S}
\fmfright{t0,t,t1,t2,t3,t4,t5,T}
\fmf{plain,width=2}{s0,s,s1,s2}
\fmf{dots}{s2,s3}
\fmf{plain,width=2}{s3,s4,s5,S}
\fmf{plain,width=2}{t0,t,t1,t2}
\fmf{dots}{t2,t3}
\fmf{plain,width=2}{t3,t4,t5,T}
\fmf{wiggly}{s,v}
\fmf{wiggly}{s1,v}
\fmf{wiggly,tension=2}{s4,v}
\fmf{wiggly}{t,v}
\fmf{wiggly}{t1,v}
\fmf{wiggly,tension=2}{t4,v}
\fmfv{d.sh=circle,d.f=empty,d.si=20}{v}
\fmf{wiggly}{s5,t5}
\fmflabel{$0$}{s0}
\fmflabel{$0$}{t0}
\fmflabel{$S$}{S}
\fmflabel{$T$}{T}
\end{fmfgraph*}}}
\]
and analytically
\begin{equation}
\Gamma(S,T)=1+\int_0^Sds\int_0^Tdt\,
\Gamma(s,t)\frac{\lambda}{4\pi^2 [(s-t)^2+L^2]}.
\label{sd}
\end{equation}
The static potential is extracted from $\Gamma(S,T)$ as $
V(L)=-\lim_{T\rightarrow\infty}\frac{1}{T}\ln\Gamma(T,T)$.  We note
here that, generally, the sum of ladder diagrams will not produce a
gauge invariant result.  However, the few leading terms (the two terms
in (\ref{effch})) are independent of the choice of gauge.  

The kernel in (\ref{sd}) obeys the differential equation
\[
\frac{\partial ^2\Gamma(S,T)}{\partial S\,\partial T}=
\frac{\lambda}{4\pi^2\left[ (S-T)^2+L^2\right]}\Gamma(S,T)
\]
and the boundary conditions $\Gamma(0,T)=1=\Gamma(S,0)$.  This equation is
separable in the variables $x=(S-T)/L$ and $y=(S+T)/L$. The Laplace
transform of $\Gamma$ can be expressed in terms of 
eigenfunctions of the Schr\"odinger equation
\begin{equation}
\left[-\frac{d^2}{dx^2}-\frac{\lambda}{4\pi^2(x^2+1)}
\right]\psi_n(x)=
-\frac{\Omega_n^2}{4}\psi_n(x).
\label{schro}
\end{equation}
Explicitly,
\begin{equation}\label{spec}
\int_{|x|}^\infty dy\,\e^{-py}\,\Gamma(x,y)=2\sum_n
\frac{\bar{\psi}_n(0)\psi_n(x)}{p^2-\Omega^2_n/4}.
\end{equation}
The asymptotic behavior of $\Gamma(x,y)$ at large $y$ is determined by
the rightmost singularity of its Laplace transform in the complex $p$
plane. The right hand side of \rf{spec} has poles at the eigenvalues
of the Schr\"odinger equation. The pole with the largest real
component is associated with the ground state energy.  Consequently,
the sum of ladder diagrams grows exponentially:
$\Gamma(0,y)\sim\exp(\Omega_0 y/2) =\exp(\Omega_0T/L)$.  Thus, the
Coulomb charge is given by the ground state eigenvalue
$\a(\l)=\Omega_0(\l)$.
 
It is straightforward to find $\a(\lambda)$ in both the large
and small $\l$ limits.  If $\lambda$ is small, it is necessary to
rescale $x$ by $\l$, so that $\chi(z)=\psi(z/\lambda)$.  In that case,
the limit of \rf{schro} is
\[
-\chi''(z)-\frac{1}{4\pi}\delta(z)\chi(z)=-\frac{\a^2}{4\lambda^2}\chi(z).
\]
Using the lowest eigenvalue of the delta-function potential,
we reproduce the leading behavior, $
\alpha(\l)=\l/4\pi+\cdots$.   A straightforward perturbative 
computation produces the next order, 
\begin{equation}
\alpha(\l)=\frac{\l}{4\pi}-\frac{\l^2}{8\pi^3}\,
\ln\frac{1}{\lambda}+\cdots,
\label{effch}\end{equation}
including the logarithm.  There is also a (smaller) term of order
$\l^2$ which, to this order, has a gauge dependent coefficient. In any
case, infrared divergences make it impossible to compute the term of
order $\l^2$ in perturbation theory.

Even though the result is not independent of the choice of gauge, it is
interesting to study the large $\l$ behavior of the potential
arising in the infinite sum of ladder diagrams.  If $\l$ is large, the
potential in (\ref{schro}) can be expanded in $x$ about $x=0$.  The
wavefunction in the quadratic approximation is a Gaussian
\begin{equation}
\psi(x)\propto e^{-\sqrt{\lambda}x^2/4\pi},
\end{equation}
whose ground state energy is $\Omega^2=\lambda/\pi^2+\cdots$.  This
approximation is consistent; the expectation value of corrections to
the quadratic approximation for the ground state energy is of higher
order in $1/\sqrt{\lambda}$.  This yields the strong coupling behavior
\begin{equation}
\alpha(\l)=\frac{\sqrt{\l}}{\pi}-1+O(1/\sqrt{\l}),
\label{result}
\end{equation} 
which reproduces the same power of the coupling constant as in the the
AdS/CFT computation in (\ref{adscft}), but has the wrong coefficient.
The source of the discrepancy in the coefficient is clear: The ladder
diagrams do not contain all of the important contributions at strong
coupling.

The main result of this Letter is the one-loop correction to the
static potential~(\ref{effch}). It is satisfying to see that this
correction goes in the right direction to match the prediction of the
AdS/CFT correspondence.  (The sub-leading corrections on the supergravity
side have been studied in \cite{Gre98}.)  We also found that the
$\sqrt{\l}$ dependence of the effective Coulomb charge predicted by
the AdS/CFT correspondence arises in the resummed perturbation theory
and that $1/\sqrt{\l}$ is a natural expansion parameter at strong
coupling, just as in the supergravity approach.  These results are in
good agreement with the AdS/CFT correspondence.

The infrared divergences due to soft gauge and scalar bosons in the
static potential appear to be stronger in \N supersymmetric Yang-Mills
theory than in QCD. These divergences make the Coulomb potential
non-analytic at weak coupling and lead to the breakdown of
perturbation theory beyond second order.  This behavior is similar to
what happens at finite temperature, but the infrared divergences are
milder in the case of the static potential and have a different
physical origin.

\bigskip

The authors acknowledge support of NSERC of Canada. The work of
R.J.S. was supported in part by the Danish Natural Science Research
Council.  The work of K.Z. was supported by NATO Science Fellowship,
PIms Postdoctoral Fellowship and grant 96-15-96455 for the promotion
of scientific schools.

\end{fmffile}


\begin{thebibliography}{99}

\bibitem{Mal97}
J.~Maldacena,
Adv.\ Theor.\ Math.\ Phys.\ {\bf 2}, 231 (1998),
{\tt hep-th/9711200}.

\bibitem{Gub98}
S.S.~Gubser, I.R.~Klebanov and A.M.~Polyakov,
Phys.\ Lett.\ {\bf B428}, 105 (1998),
{\tt hep-th/9802109}.

\bibitem{Wit98}
E.~Witten,
Adv.\ Theor.\ Math.\ Phys.\ {\bf 2}, 253 (1998),
{\tt hep-th/9802150}.

\bibitem{Aha99}
O.~Aharony, S.S.~Gubser, J.~Maldacena, H.~Ooguri and Y.~Oz,
{\tt hep-th/9905111}, and references therein.

\bibitem{amp}A. M. Polyakov, Int. J. Mod. Phys. {\bf A14}, 645 (1998), 
{\tt hep-th/9809057}; Nucl. Phys. Proc. Suppl. {\bf 68}, 1 (1998),
{\tt hep-th/9711002}; ``{\it Gauge Fields and Strings}'', Harwood Press
(1987).

\bibitem{8}
S.~Lee, S.~Minwalla, M.~Rangamani and N.~Seiberg,
Adv. Theor. Math. Phys. {\bf 2}, 697 (1998), {\tt hep-th/9806074}; J.
Erickson, G. W. Semenoff and K. Zarembo, Phys. Lett. {\bf B466}, 239
(1999), {\tt hep-th/9906211}.

\bibitem{dine}M. Dine and J. Gray, {\tt hep-th/9909020};
%
E.~D'Hoker, D.Z.~Freedman and W.~Skiba,
Phys. Rev. {\bf D59}, 045008 (1999)
{\tt hep-th/9807098};
F.~Gonzalez-Rey, B.~Kulik and I.Y.~Park,
{\tt hep-th/9903094}.
\bibitem{thooft}G. 't~Hooft, Comm. Math. Phys. {\bf 86}, 449 (1982).

\bibitem{bipz}E. Brezin, C. Itzykson, G. Parisi and J. Zuber, Comm. Math. 
Phys. {\bf 59}, 35 (1979).

\bibitem{Gub96}
S.S.~Gubser, I.R.~Klebanov and A.W.~Peet,
Phys.\ Rev.\ {\bf D54}, 3915 (1996)
{\tt hep-th/9602135}.

\bibitem{Gub98'}
S.S.~Gubser, I.R.~Klebanov and A.A.~Tseytlin,
Nucl.\ Phys.\ {\bf B534}, 202 (1998)
{\tt hep-th/9805156}; 
T.~Harmark and N.A.~Obers,
{\tt hep-th/9910036}.

\bibitem{Fot98}
A.~Fotopoulos and T.R.~Taylor,
Phys.\ Rev.\ {\bf D59}, 061701 (1999)
{\tt hep-th/9811224};
M.A.~Vazquez-Mozo,
Phys.\ Rev.\ {\bf D60}, 106010 (1999)
{\tt hep-th/9905030};
C.~Kim and S.~Rey,
{\tt hep-th/9905205};
A.~Nieto and M.H.~Tytgat,
{\tt hep-th/9906147}.

\bibitem{Mal98}
J.~Maldacena,
Phys.\ Rev.\ Lett.\ {\bf 80}, 4859 (1998)
{\tt hep-th/9803002};
S.~Rey and J.~Yee,
{\tt hep-th/9803001}.


\bibitem{Fis77}
W.~Fischler,
Nucl.\ Phys.\ {\bf B129}, 157 (1977).

\bibitem{App78}
T.~Appelquist, M.~Dine and I.J.~Muzinich,
Phys.\ Rev.\ {\bf D17}, 2074 (1978).

\bibitem{Gre98}
J.~Greensite and P.~Olesen,
JHEP {\bf 08}, 009 (1998),
{\tt hep-th/9806235};
S.~Forste, D.~Ghoshal and S.~Theisen,
JHEP {\bf 08}, 013 (1999),
{\tt hep-th/9903042};
S.~Naik,
{\tt hep-th/9904147};
J.~Sonnenschein,
{\tt hep-th/9910089}; Y. Kinar, E. Schreiber, J. Sonnenschein and N. Weiss,
in preparation.

\end{thebibliography}
\end{document}